# Comment on "Unified treatment for the evaluation of arbitrary multielectron multicenter molecular integrals over Slater-type orbitals with noninteger principal quantum numbers"

I.I.Guseinov

*Department of Physics, Faculty of Arts and Sciences, Onsekiz Mart University, Çanakkale, Turkey*

**Abstract**

Özdoğan (Int. J. Quantum Chem., 92 (2003) 419) published formulas for evaluating the multielectron multicenter molecular integrals over Slater-type orbitals (STOs). It is demonstrated that the formulas presented in this work are not original and they can easily be derived by means of a simple algebra from the relationship of our published papers (I.I.Guseinov, J.Mol.Struct.(Theochem), 417(1997)117; J.Mol.Struct.(Theochem), 593 (2002) 65; I.I.Guseinov,B.A.Mamedov,F.Öner,S.Hüseyin, J.Mol.Struct.(Theochem), 545(2001)265; I.I.Guseinov,B.A.Mamedov, J.Mol.Model., 8(2002)272; Theor.Chem.Acc., 108(2002)21).

## I. Introduction

It is well known that the variational method for improving the Hartree-Fock (HF) solution in which the interelectronic coordinates are explicitly included in the wave functions was first introduced by Hylleraas [1,2]. The solutions obtained by the Hylleraas method converge to the exact solution of Schrödinger equation with any desired degree of accuracy if a sufficient number of terms are included. However, it is difficult to evaluate the multicenter integrals that arise in Hylleraas theory. The Hylleraas method developed first by James and Coolidge [3] has been used for determining the ground state energy of the $H_2$ molecule [4,5] and still valid for two-, three- and four-electron atomic and molecular systems [6-15]. In Refs. [16] and [17], we established the general formulae for the multicenter-multielectron integrals over integer and noninteger $n$ STOs (ISTOs and NISTOs) with an arbitrary s-electron operator $F^{(s)}$ $(s=1,2,3,...)$ using complete orthonormal sets of Lambda and $\Psi^\alpha$ $(\alpha = 1, 0, -1, -2, ...)$

exponential type orbitals, respectively. These integrals arise in the solution of multielectron atomic and molecular problems when a Hylleraas approximation in the Hartree-Fock theory is employed.

In a recent paper in this journal, Özdoğan [18] published the formulas for the evaluating of multielectron multicenter integrals over NISTOs. In this Comment we demonstrate that the results published by this author in multielectron multicenter integrals are not original and can be derived from the formulas published in our papers [16, 19-22]. It should be noted that the work of Özdoğan [18] is on the use of formulas presented in Refs.[23-16], which are obtained from our papers by changing the summation indices or application of a simple algebra (see Comment [28-30]).

The published in Ref.[18] formulas in multielectron multicenter molecular integrals can be obtained from our papers [16,19-22] basic formulas of which are presented in the following sections:

**2.** Expansion of NISTOs through ISTOs with the same center (see Eqs. (6)-(10) of Ref.[20] and Eqs. (3)-(6) of [16] for $\vec{R}_{ab} = 0$):

$$\chi_{nlm}(\zeta,\vec{r}) = \lim_{N \to \infty} \sum_{n'=l+1}^{N} V_{nl,n'l}^{N}(t) \chi_{n'lm}(\zeta',\vec{r}), \tag{1}$$

where $t = (\zeta - \zeta')/(\zeta + \zeta')$ and

$$V_{nl,n'l}^{N}(t) = \sum_{n'=l+1}^{N} \sqrt{a_{nn'}}\, \Omega_{n'n''}^{l}(N)(1+t)^{n+\frac{1}{2}}(1-t)^{n''+\frac{1}{2}}, \tag{2}$$

$$a_{nn'} = \begin{cases} F_{2n'}(n+n')/F_{n+n'}(2n) & \text{for } n \geq n' \\ F_{2n}(n+n')/F_{n+n'}(2n') & \text{for } n \leq n' \end{cases}, \tag{3}$$

$$\Omega_{nn'}^{l}(N) = \sum_{n''=\max(n,n')}^{N} \omega_{n''n}^{l} \omega_{n''n'}^{l}, \tag{4}$$

$$\omega_{n'n}^{l} = (-1)^{n+l+1}\left[F_{n-l-1}(n'-l-1)F_{n-l-1}(2n)F_{n+l+1}(n'+l+1)\right]^{\frac{1}{2}}. \tag{5}$$

**3.** Unified expression of arbitrary in multielectron multicenter integrals over NISTOs (see Eqs. (11), (25) and (26) of [16], and also Eqs. (19)-(21) of [19]):

$$I^{(s)}_{p_1^* p_1^{\prime*}, p_2^* p_2^{\prime*}, \ldots, p_s^* p_s^{\prime*}}(\zeta_1, \zeta_1', \vec{R}_{ca}, 0; \zeta_2, \zeta_2', \vec{R}_{db}, \vec{R}_{ba}; \ldots; \zeta_s, \zeta_s', \vec{R}_{fe}, \vec{R}_{ea})$$
$$= \int F^{(s)} \rho^*_{p_1^* p_1^{\prime*}}(\zeta_1, \vec{r}_{a1}; \zeta_1', \vec{r}_{c1}) \rho_{p_2^* p_2^{\prime*}}(\zeta_2, \vec{r}_{b2}; \zeta_2', \vec{r}_{d2}) \ldots \rho_{p_s^* p_s^{\prime*}}(\zeta_s, \vec{r}_{es}; \zeta_s', \vec{r}_{fs}) dV_1 dV_2 \ldots dV_s, \tag{6}$$

$$\rho_{p^* p^{\prime*}}(\zeta, \vec{r}_g; \zeta', \vec{r}_h) = \chi_{p^*}(\zeta, \vec{r}_g)\chi^*_{p^{\prime*}}(\zeta', \vec{r}_h),$$

$$I^{(s)}_{p_1^* p_1^{\prime*}, p_2^* p_2^{\prime*}, \ldots, p_s^* p_s^{\prime*}}(\zeta_1, \zeta_1', \vec{R}_{ca}, 0; \zeta_2, \zeta_2', \vec{R}_{db}, \vec{R}_{ba}; \ldots; \zeta_s, \zeta_s', \vec{R}_{fe}, \vec{R}_{ea}) \tag{7}$$

$$= \lim_{N_1,N_2,...,N_s \to \infty} \sum_{\mu_1=1}^{N_1}\sum_{\nu_1=0}^{\mu_1-1}\sum_{\sigma_1=-\nu_1}^{\nu_1} W^{*N_1}_{p_1^* p_1'^* q_1}(\zeta_1,\zeta_1',z_1;\vec{R}_{ca},0) \sum_{\mu_2=1}^{N_2}\sum_{\nu_2=0}^{\mu_2-1}\sum_{\sigma_2=-\nu_2}^{\nu_2} W^{N_2}_{p_2^* p_2'^* q_2}(\zeta_2,\zeta_2',z_2;\vec{R}_{db},\vec{R}_{ba})...$$

$$\sum_{\mu_s=1}^{N_s}\sum_{\nu_s=0}^{\mu_s-1}\sum_{\sigma_s=-\nu_s}^{\nu_s} W^{N_s}_{p_s^* p_s'^* q_s}(\zeta_s,\zeta_s';\vec{R}_{fe},\vec{R}_{ea}) J^{(s)}_{q_1 q_2 ... q_s}(z_1,z_2,...,z_s). \qquad (8)$$

The quantities $J^{(s)}$ occurring in Eq.(8) are the one-center $s$-electron basic integrals with ISTO's which were defined and evaluated in Ref.[19] for the special cases of one- and two-electron operators $F^{(1)} = \frac{1}{r_{b1}}$ and $F^{(2)} = \frac{1}{r_{21}}$ appearing in the HFR equations for molecules.

## 4. Application to one-electron molecular integrals over NISTOs

### 4.1. Two-center overlap integrals

In order to derive the expression for evaluation of overlap integrals over NISTOs in terms of overlap integrals with ISTOs we should use the expansion formulas (1). Then, we get Eq.(17) of Ref.[18]. The analytic formulae for two-center overlap integrals of NISTOs and ISTOs have been published in our papers (see e.g., Ref.[21]). The published by Özdoğan formulas for two-center overlap integrals are obtained from our results by changing the summation indices (see Comment [28]).

### 4.2 Two-center nuclear attraction integrals $U^{(A,B)}$

Eq.(20) of Ref.[18] for integrals $U^{(A,B)}$ is obtained with the help of one-center expansion formula (1) of this Comment. It should be noted that the two-center nuclear attraction integrals $U^{(A)}$ and $U^{(B)}$ are reduced to the two-center overlap integrals (see Comment [29]).

### 4.2 Three-center nuclear attraction integrals

Eq.(22) of Ref.[18] for three-center nuclear attraction integrals is also obtained by taking into account the one-center expansion formula (1). It should be noted that the three-center nuclear attraction integrals over NISTOs and ISTOs have been published by us in Ref.[22].

## 5. Application to two-electron molecular integrals over NISTOs

Eqs.(23)-(31) of published by Özdoğan paper [18] can be easily derived from formulas (6) and (8) of this Comment for the special case of two-electron operator $F^{(2)} = \frac{1}{r_{21}}$. The results for the evaluation of multicenter two-electron integrals with ISTOs and NISTOs have been given in Ref.[16].

As can be seen from the formulas given in this Comment and in Ref.[18], all of the formulas published by Özdoğan for the evaluation of multielectron multicenter molecular integrals over NISTOs and ISTOs are available in our papers or they can easily be derived by means of a simple algebra. It should be noted that the formulas presented in [23-27] for the calculation of two-center overlap and nuclear attraction integrals over STOs in nonlined-up and lined-up coordinate systems can also be derived, as shown in Comments [28-30], from the use of established in our papers formulas.

**References**
1. E.A.Hylleraas,Z.Physik.,48(1928)469.
2. a) E.A.Hylleraas,Z.Physik.,54(1929)347;   b) E.A.Hylleraas,Z.Physik.,60(1930)624;   c) E.A.Hylleraas,Z.Physik.,65(1930)209.
3. H.M.James,A.S.Coolidge,J.Chem.Phys.,1(1933)825.
4. W.Kolos,C.C.J.Roothaan, Rev.Mod.Phys.,32(1960)205;   b) W.Kolos, C.C.J.Roothaan, Rev.Mod.Phys.,32(1960)219.
5. W.Kolos,L.Wolniewicz,J.Chem.Phys.,41(1964)3663;   W.Kolos,L.Wolniewicz, J. Chem. Phys.,49(1968)404.
6. Y.Öhrn,J.Nording, J. Chem. Phys.,39(1963)1864.
7. S.Larson , Phys. Rev.A,169(1968)49.
8. J.F.Perkins, Phys. Rev.A,13(1976)915.
9. D.M.Fromm,R.N.Hill, Phys. Rev. A,36(1987)1013.
10. E.Remiddi, Phys. Rev. A,44(1991)5492.
11. F.W.King, Phys. Rev. A,44(1991)7108.
12. F.W.King, J. Chem. Phys.,99(1993) 3622.
13. A.Luchow,H.Kleindienst, Int. J. Quantum Chem.,45(1993)445.
14. I.Porras,F.W. King, Phys. Rev. A,49(1994)1637.
15. H.Kleindienst,A.Luchow ,Phys. Rev. A,51(1995)5019.
16. I.I.Guseinov, J.Mol.Struct.(Theochem),593(2002)65.
17. I.I.Guseinov, J.Mol.Mod.,9(2003)190.
18. T.Özdoğan, Int.J.Quantum Chem.,92(2003)419.
19. I.I.Guseinov, J.Mol.Struct.(Theochem),417(1997)117.
20. I.I.Guseinov,B.A.Mamedov,F.Öner,S.Hüseyin, J.Mol.Struct.(Theochem),545(2001)265.
21. I.I.Guseinov,B.A.Mamedov, J.Mol.Mod.,8(2002)272.
22. I.I.Guseinov,B.A.Mamedov, Theor.Chem.Acc.,108(2002)21.
23. T.Özdoğan,M.Orbay, Int.J.Quantum Chem.,87(2002)15.
24. T.Özdoğan,M.Orbay,S.Gümüş, Commun.Theor.Phys.,37(2002)711.
25. T.Özdoğan,S.Gümüş,M.Kara, J.Math.Chem.,33(2003)181.
26. S.Gümüş,T.Özdoğan, Commun.Theor.Phys.,39(2003)701.
27. E.Öztekin,M.Yavuz,Ş.Atalay, J.Mol.Struct.(Theochem),544(2001)69.

28. I.I.Guseinov, Int.J.Quantum Chem.,91(2003)62.
29. I.I.Guseinov, J.Math.Chem.,35(2004)327.
30. I.I.Guseinov, J.Mol.Struct.(Theochem), 638(2003)235.